\documentclass[conference]{IEEEtran}
\IEEEoverridecommandlockouts
\usepackage{cite}
\usepackage{amsmath,amssymb,amsfonts}
\usepackage{algorithmic}
\usepackage{graphicx}
\usepackage{textcomp}
\usepackage{xcolor}
\usepackage{multirow}
\usepackage[a4paper, total={184mm,239mm}]{geometry}
\usepackage{booktabs}

\def\BibTeX{{\rm B\kern-.05em{\sc i\kern-.025em b}\kern-.08em
    T\kern-.1667em\lower.7ex\hbox{E}\kern-.125emX}}
\usepackage{todonotes}
\renewcommand{\baselinestretch}{0.975}
\begin{document}

\title{Model Predictive Current Control with Harmonic Correction for Single-Phase AC-DC EV Charging
\thanks{
This work was supported by EU MSCA Project “COALESCE” under Grant Number 101130739 and the Sustainable Energy Authority of Ireland under Grant Number 24/RDD/1170.
}}

\author{
Changhong Li,
Bharathkumar Hegde,
Biswajit Basu,
Shreejith Shanker \\ 
Reconfigurable Computing Systems Lab, Electronic \& Electrical Engineering\\
Trinity College Dublin, Ireland\\
Email: \{lic9, shripa, basub, shreejith.shanker\}@tcd.ie
}

\maketitle

\begin{abstract}
The increasing integration of Electric Vehicles (EVs) has imposed a growing harmonic challenge on the power grid.
For AC/DC Power Factor Correction (PFC) in single-phase On-Board Chargers (OBCs), Model Predictive Current Control (MPCC) improves the current quality by predicting and tracking the inductor current.
However, finite control set MPCC selects switching states, resulting in discrete control actions and a limited optimisation space.
Moreover, the MPCC's cost function based on instantaneous current tracking error has limited capability to compensate for low-order harmonic disturbances induced by dead time, control delay, and model parameter mismatch. 
This paper proposes a duty cycle predictive MPCC incorporating a real-time harmonic estimation reference. 
The proposed method dynamically estimates the low-order harmonic components of the input current and corrects the MPCC reference current, enabling continuous duty cycle control and targeted suppression of dominant low-order harmonics.
Simulation results on a single-phase OBC demonstrate that the proposed duty cycle predictive MPCC reduces the steady current $THD_i$ from 11.47\% to 6.10\% compared with the switching state predictive MPCC. With the harmonic reference, the $THD_i$ is further reduced to 2.85\%.
\end{abstract}


\section{Introduction}\label{introduction}
The rapid penetration of EVs is increasing the interaction between transportation electrification and distribution power systems.
EV charging facilities are typically interfaced with the grid through power electronic converters, which behave as nonlinear loads and may introduce current harmonics, voltage distortion, power factor degradation, additional losses, and accelerated ageing of grid equipment~\cite{srivastava2023electric, harish2022review}.
International standard IEC-61000-4-7 have been established to limit harmonics that degrade power quality and the standard procedures to conduct the measurement~\cite{iec-61000-4-7}.
To satisfy these standards, the suppression of harmonics at the source of injection is crucial for EV charging systems. 

A typical two-stage OBC first rectifies the AC input and performs PFC, and then regulates the battery charging voltage and current through a downstream DC/DC stage \cite{dar2024board, yuan2021review}.
In conventional PFC control, Proportional Resonant (PR) control is commonly adopted for the inner loop because it provides high gain at the fundamental frequency.
However, its performance is sensitive to grid frequency deviation, distorted input voltage, and switching nonidealities and may inject more harmonic distortion.


MPCC has become an attractive alternative for PFC converters.
MPCC predicts the future inductor current under candidate control actions and selects the optimal action according to a current tracking objective, which provides a direct way to handle converter nonlinearities~\cite{ko2023model}.
However, conventional finite control set MPCC directly selects binary switching states.
This discrete control action restricts the optimisation space and may cause larger current ripple and variable switching.

High-frequency switching actions in single-phase PFC converters, together with non-ideal device characteristics such as dead time, introduce harmonics into the current, among which low-order odd harmonics are the most prominent~\cite{jeong1991analysis, yang2015harmonics}.
By embedding the harmonic estimation into reference generation, the controller can actively correct the current reference and provide selective compensation for targeted harmonic suppression.
However, this relies on fast and accurate harmonic estimation.

Conventional harmonic estimation methods, such as Fast Fourier Transform (FFT) and its variants, are widely used for frequency-domain analysis, but their accuracy can degrade under non-stationary operating conditions due to spectral leakage, aliasing, and picket-fence effects.
A series of more advanced time-frequency estimation methods have been introduced and integrated into edge devices to improve the robustness and accuracy of harmonic estimation~\cite{norman2012hybrid, gu2007estimating, tiwari2016hardware, tiwari2017fast, baraskar2022digital, li2024fast}.
Recent data-driven harmonic estimation approaches with Broad Learning System (BLS)~\cite{li2023broad} demonstrated the robustness with noise and frequency conversion, estimating the harmonic components with less than 1 power cycle offline.
In~\cite{taghvaie2025online}, the authors demonstrates that Temporal Convolutional Network (TCN) can be deployed online for real-time harmonic estimation.
However, these promising deep learning-based estimation approaches have not been integrated into the control loop to improve the power quality.
To bridge these gaps, this paper proposes a duty cycle predictive MPCC with a real-time harmonic estimation reference for the PFC stage of a single-phase OBC, in order to improve the power quality from the EV charging end.

The main contributions of this paper are as follows:
\begin{itemize}
    \item A duty cycle predictive MPCC for the single-phase OBC's PFC, reducing harmonics by replacing the direct discrete switch control with continuous duty cycle optimisation.
    
    \item A deep learning-based real-time harmonic estimation model for MPCC current reference correction, enabling targeted suppression of low-order odd harmonics.
    
    \item OBC Hardware-in-the-loop simulation results show that the proposed MPCC reduces $THD_i$ from 11.47\% to 6.10\%, while with the harmonic reference reaching 2.85\%.
\end{itemize}

\section{Related Work}\label{background}
\subsection{Model Predictive Current Control}
Typically, PFCs used in AC-DC pre-regulators of power converters consist of an outer voltage controller loop and an inner current controller~\cite{rahman2021design}. Specifically, the PR controller is widely used in the inner current controller. However, it is susceptible to voltage distortion and is sensitive to noise.
Predictive current mode control (PCMC)~\cite{zhang2003new, azazi2011dsp, chen2003predictive} generates a sinusoidal current reference with the same frequency as the input voltage internally.
With a predicted current reference, it has the advantage of being robust to voltage distortion.

Model predictive control (MPC) defines a predictive model and a cost function based on a system model. The control input is calculated by minimising the cost function~\cite{bordons2015basic, mayne2014model}.
Its current control version, MPCC, is employed for boost PFC converter's current control~\cite{bartsch2016analysis}.
The future inductor current value when the switch is turned on/off is predicted and compared with the reference value, and the switching operation of the next cycle is determined by minimising the cost function.

Recent work in~\cite{ko2023model} has shown that by using variable switching action, MPCC can reduce switching activity in areas of high current stress, thereby improving converter efficiency.
However, switch state predictive MPCC is restricted to a finite set of discrete control actions, resulting in a limited optimisation space and constraining further improvements in power quality.
Although MPCC minimises the predicted current tracking error within each sampling interval, a purely sinusoidal current reference does not explicitly account for low-order harmonic disturbances caused by the non-ideal factors like dead time.
By incorporating real-time harmonic estimation into the reference generation process, the controller can reshape the current reference with selective compensation terms and suppress the dominant low-order odd harmonics more effectively.

\subsection{Harmonic Estimation Model}
Harmonic estimation involves analysing the distorted voltage/current signals to identify the harmonic components. 
In practice, the FFT remains a fundamental and effective method for harmonic analysis.
Several advanced signal processing methods have also been explored to improve estimation performance under leakage, noise, and non-stationary conditions, including wavelet-based, Hilbert-transform-based analysis, and optimisation or model-based techniques like ESPRIT~\cite{norman2012hybrid, gu2007estimating}.
Although these conventional approaches are effective in most ideal scenarios, most of them rely on longer observation windows or stronger signal assumptions.
Recently, data-driven approaches like BLS have also been introduced for harmonic estimation and have demonstrated robust performance with noise and frequency deviation~\cite{li2023broad}.
However, it involves more computation and has higher latency.

\begin{figure}[htbp]
    \vspace{-3mm}
    \centering
    \includegraphics[width=0.45\textwidth]{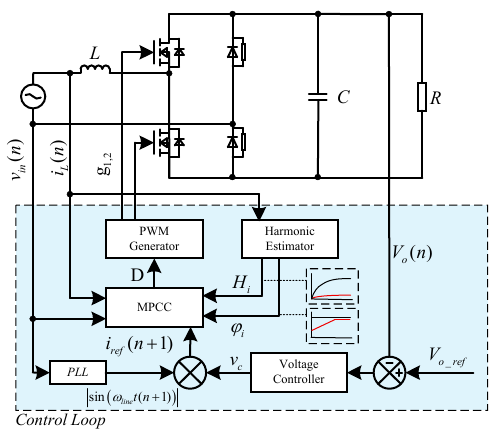}
    \caption{Proposed harmonic-estimation-assisted duty cycle predictive MPCC control loop for the single-phase AC/DC PFC converter.}
    \label{fig:framework}
    \vspace{-4mm}
\end{figure}

A series of algorithms was offloaded to edge devices for online harmonic estimation.
The Discrete Wavelet Packet Transform (DWPT) \cite{tiwari2016hardware} and its variant Undecimated Wavelet Packet Transform (UWPT) \cite{tiwari2017fast} are offloaded to FPGAs for real-time harmonics estimation.
In~\cite{baraskar2022digital}, the authors show that an FPGA-based DWPT implementation improved hardware efficiency by using MAC-based wavelet filters and level-wise downsampling.
In \cite{li2024fast}, the authors show the comparative advantage of offloading FFT-based harmonic analysis to FPGAs over MCUs, with 33$\times$ performance improvement reported for the FPGA implementation.
Other algorithm like Gauss-Newton Optimisation (GNO) was implemented on DSPs for real-time harmonic estimation under power system frequency deviations~\cite{genccol2023efficient}.
In~\cite{taghvaie2025online}, the deep learning based TCN model for harmonic estimation is deployed online to estimate the harmonics in a single-drive system with a 4-kW. However, these estimations were not incorporated into real-time control.
\section{Methodology and Design} \label{methodology}

Fig.~\ref{fig:framework} illustrates the proposed harmonic estimation-assisted MPCC control framework.
The main circuit adopts a single-phase AC/DC PFC converter structure, which is used for AC input current shaping and DC link voltage regulation.
In the control stage, the outer voltage loop generates the current amplitude command according to the error between the output voltage \(V_o(n)\) and the reference voltage \(V_\text{o,ref}\), while the PLL provides the phase information synchronised with the input voltage to construct the fundamental current reference.
The harmonic estimation module performs online estimation of the low-order harmonic components in the input current \(i_L(n)\), yielding the corresponding harmonic amplitudes \(H_i\) and phases \(\varphi_i\), which are then used to correct the current reference \(i_\text{ref}(n+1)\) for the next sampling period.
Based on the converter predictive model, the MPCC controller computes the optimal duty cycle \(D\), which is applied to the power switches through the PWM generator to achieve closed-loop input current tracking and low-order harmonic suppression.
The following sections present the harmonic estimation model and the MPCC current control algorithm with harmonic reference correction, respectively.

\subsection{BLS-based Real-time Harmonic Estimation Model}
For the construction of the harmonic estimation model, we follow the structure of the BLS used for harmonic estimation~\cite{li2023broad}.
The raw current waveform containing harmonic components is used as the model input.
At a sampling rate of 4000 Hz, one fundamental cycle contains 80 samples.
The backbone network inherits the BLS architecture, consisting of broad feature layers and enhancement layers.
A high-dimensional feature space is constructed through a large number of random mappings for feature extraction. 
The broad feature layers and enhancement layers serve as the feature extractor, while multiple task-specific output heads are trained by closed-form regression rather than backpropagation to perform harmonic estimation. 
This design facilitates fast training and adaptive deployment on edge devices.

Different from existing BLS-based harmonic estimation models, the original BLS model is mainly used for estimating the amplitudes of different harmonic orders.
However, amplitude information alone is insufficient for harmonic compensation.
The proposed model further estimates the relative phase of each harmonic component with respect to the fundamental component.
The phase estimation typically relies on orthogonal projection and half-wave symmetry over a complete fundamental cycle.
Therefore, a full-cycle input is adopted instead of a half-cycle or shorter input window.

Although the BLS-based harmonic estimation model improves representation capability through wide feature layers, it also increases the computational burden, which makes real-time edge deployment challenging.
This characteristic is well matched with the pipelined parallel computing paradigm on an FPGA, where low-latency inference can be achieved through spatial parallelism.
To further improve real-time performance, 3-bit quantisation is applied to the weights and activations in the feature layers and enhancement layers.
The input and output layers are quantised with 8 bits, and the network is implemented in a fully unrolled manner.
In addition, sparse weights are introduced in selected feature layers to further reduce the model size and inference latency.
The resulting quantised neural network is finally deployed on the FPGA side through FINN~\cite{blott2018finn, umuroglu2017finn} and is invoked by the MPCC controller for real-time harmonic estimation.
\vspace{-1mm}

\subsection{MPCC with Harmonic Estimations}
The proposed controller optimises the duty cycle input for the PWM in the single-phase AC/DC PFC converter using harmonic estimation-assisted MPCC control. 
Compared with a conventional PR current controller, the proposed method does not rely only on a linear resonant compensator to reduce the current tracking error at the fundamental frequency.
Instead, it uses the discrete converter model to calculate the duty cycle $D_{\text{opt}}$ to minimise predicted harmonics, error between the predicted current and the current reference, and excessive current oscillation. 

To minimise the influence of low-order harmonics on the input current, the estimated harmonic components are subtracted from the predicted current reference, thereby generating a harmonic-compensated reference signal.
The current reference prediction is expressed as:
\begin{equation}
    i_{\text{ref}}(n+1) = v_c|\sin(\omega_{\text{line}}t(n+1))|
    \label{eq:i_ref}
\end{equation}
where, $v_c$ is the output of the voltage controller and $\omega_{\text{line}}$ is the frequency of the input voltage. $v_c$ represents the magnitude, and $|\sin(\omega_{\text{line}}t(n+1))|$ is a predicted sine wave that determines the shape of the current reference.

The amplitudes and phases of the low order harmonics are estimated online using the proposed BLS-based estimation model explained previously.
The controller reconstructs the dominant low order harmonics from the estimated amplitudes and phases. 
The PLL phase increment is first calculated as
\begin{equation}
\begin{aligned}
\Delta \theta(n)
&=
\operatorname{atan2}
\Big(
\sin\big(\theta_{\text{pll}}(n)-\theta_\text{pll}(n-1)\big), \\
&\qquad
\cos\big(\theta_{\text{pll}}(n)-\theta_\text{pll}(n-1)\big)
\Big).
\end{aligned}
\label{eq:delta_theta}
\end{equation}
and the instantaneous angular frequency is estimated by
\begin{equation}
\omega_\text{inst}(n)
=
\frac{\Delta \theta(n)}{T_s}.
\end{equation}
A leading control phase is then used to compensate for the digital control delay:
\begin{equation}
\theta_c(n+1)
=
\theta_\text{pll}(n)+\omega_\text{inst}(n)T_s,
\end{equation}
Based on the estimated harmonic information, the 3rd, 5th, and 7th harmonic components are reconstructed as
\begin{equation}
i_h(n+1)
=
\sum_{i\in\{3,5,7\}}
A_i
\sin\left(i\theta_c(n+1)+\varphi_i\right),
\end{equation}
where \(A_i\) and \(\varphi_i\) denote the estimated amplitude, and estimated phase of the \(i\)th harmonic, respectively.

The corrected current reference is then evaluated as follows:
\begin{equation}
i_\text{ref,c}(n+1)
=
i_\text{ref}(n+1)-i_h(n+1)
\label{eq:i_refc}
\end{equation}

The predicted current $i_L(n+1)$ can be modelled for a duty cycle $D$ as follows:
\begin{equation}
    i_L(n+1) = i_L(n) - \frac{|v_{\text{in}}| - \left(1-D\right) V_o}{L}T_s
    \label{eq:i_L}
\end{equation}
where $i_L(n)$, $v_{\text{in}}$, and $V_o$ are the input current, input voltage and output voltage, respectively. 

The primary objective of the proposed MPCC is to minimise the error between the harmonic corrected reference current and the predicted current, which is defined as 
\begin{equation}
J_c = i_{\text{ref,c}}(n+1) - i_L(n+1)
\end{equation}
By substituting Eq.(\ref{eq:i_refc}) and Eq.(\ref{eq:i_L}) in the above cost function, the optimal duty cycle $D_c$ to achieve zero cost  can be evaluated as 
\begin{equation}
D_{c}
=
S\cdot\left(i_{\text{ref,c}}(n+1)-i_L(n)\right)\cdot
\frac{L}{V_o(n)T_s} - \frac{|v_{\text{in}}|}{V_o}+1
\label{eq:D_c}
\end{equation}
where $S = \operatorname{sign}\left(v_\text{in}(n)\right)$ represents the sign of the current input cycle.

Further, the controller minimises the following damping cost function to avoid excessive current oscillations, 
\begin{equation}
    J_d = i_{\text{ref,c}}(n+1) - (1-D)i_L(n+1)
\end{equation}
This cost function penalises the difference between the reference current and the average inductor current, thereby discouraging large current excursions and improving transient damping.
The optimal duty cycle to minimise the damping cost function to zero is evaluated as
\begin{equation}
D_{d}
=
\frac{\left(i_L(n)-i_\text{ref,c}(n+1)\right)}{i_L(n)}
\label{eq:D_d}
\end{equation}

\begin{figure}[htbp]
    \vspace{-3mm}
    \centering
    \includegraphics[width=\linewidth]{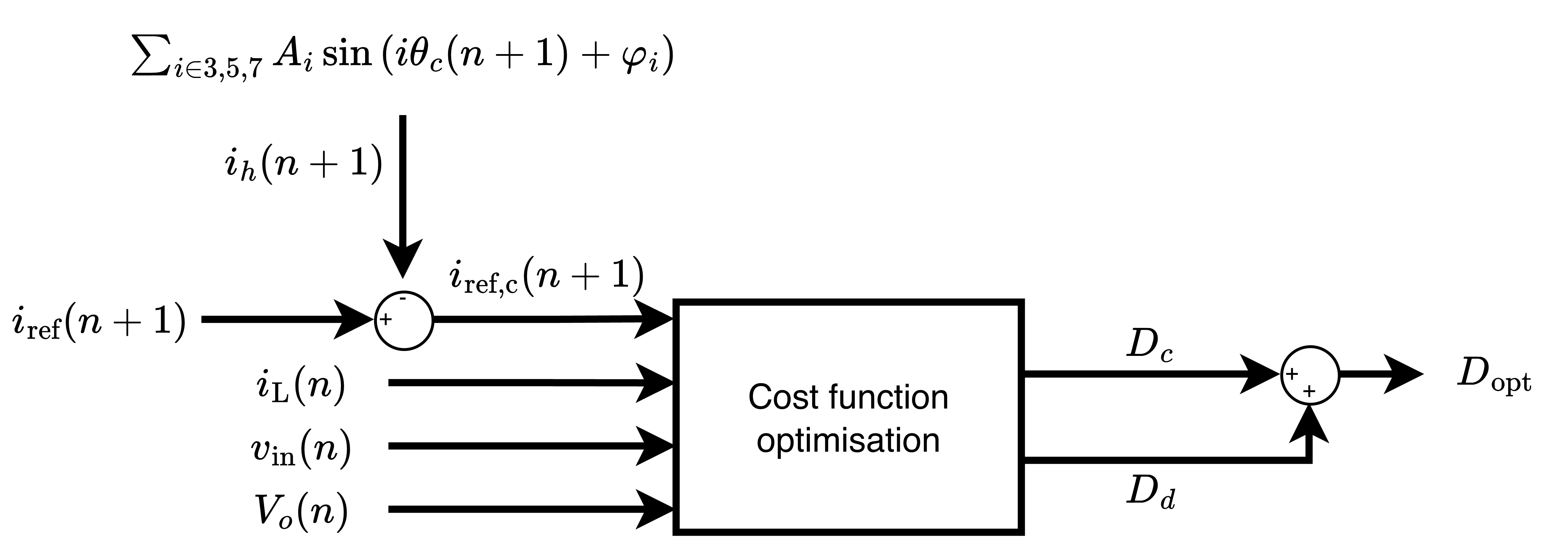}
    \caption{Proposed MPCC optimisation block with harmonic predictions}
    \label{fig:mpcc_block}
    \vspace{-2mm}
\end{figure}

The overall cost function for the proposed MPCC is defined as
\begin{equation}
J = J_c + J_d
\label{eq:J_overall}
\end{equation}
The optimal duty cycle to minimise the overall cost function $J$ to zero is evaluated by combining Eq.~(\ref{eq:D_c}) and Eq.~(\ref{eq:D_d}) as follows
\begin{equation}
D_\text{opt}
=
D_c + D_d
\end{equation}
The evaluated duty cycle $D_\text{opt}$ is saturated to limit its range within $[0,1]$ and applied to the PWM generator.
The block diagram for the proposed MPCC controller for evaluating the optimal duty cycle $D_\text{opt}$ is illustrated in Fig.~\ref{fig:mpcc_block}.

The MPCC tracks a reference that has already been corrected according to the estimated harmonic distortion.
This provides a selective feedforward compensation path for the dominant 3rd, 5th, and 7th harmonics, while retaining the model-based duty cycle prediction of MPCC.
The proposed controller, therefore, combines the regular PWM operation of duty cycle control, the predictive current tracking capability of MPCC, and the targeted harmonic suppression capability provided by the online harmonic estimator.
This duty cycle formulation is different from the finite control set MPCC used for direct switch control.
By contrast, the proposed controller outputs a continuous duty cycle for fixed frequency PWM modulation.
Therefore, the control action has a finer resolution, the current waveform is smoother, and the switching behaviour is more regular.

\section{Experimental Results}\label{results}

\subsection{Experimental Setup}
We conducted two case studies to evaluate the proposed framework.
The first case study benchmarks the accuracy and real-time performance of the harmonic estimation model under synthetic signals offline.
The dataset follows scenario A1 from the BLS evaluation for fair comparison with existing harmonic estimation approaches, especially BLS~\cite{li2023broad}.
The fundamental frequency \(f_0\) is set to \(50\,\mathrm{Hz}\) and allowed to vary within \(\pm 0.5\%\), while the amplitudes of the first, third, fifth, and seventh harmonics fluctuate within \(\pm 1\%\).
Gaussian white noise with a signal-to-noise ratio (SNR) of \(26\,\mathrm{dB}\) is added.

The second case study evaluates the closed-loop PFC performance when the harmonic estimation reference is integrated into the proposed MPCC.
The main circuit simulation is conducted in MATLAB R2025b, and the OBC simulation model is recreated from an open source OBC model~\cite{giroux2026evobc}.
The FPGA-based control unit is connected to the MATLAB simulation through the MATLAB FPGA communication toolbox, enabling hardware-in-the-loop evaluation of the controller with the simulated OBC plant.
It compares the proposed harmonic estimation-assisted MPCC with other conventional approaches.

The simulation, model training and hardware compilation are conducted on a workstation equipped with an NVIDIA RTX A4000 GPU and an Intel Core Ultra 7 265K CPU.
The software environment uses Python~3.9, PyTorch~2.5.1, CUDA~12.8, and Vivado~2022.2.
The accelerator is generated for the AMD ZU7EV FPGA on the ZCU104 development kit.

\subsection{Synthetic Signal Harmonic Offline Estimation}
We first evaluated our harmonic estimation offline before integrating to the MPCC control as a real-time reference.
The harmonic estimation relative error in the simulation scenario A1 is summarised in Table~\ref{tab:rel_error_methods}. 
Performance of competing models is recreated from ~\cite{li2023broad}, and compared to the proposed harmonic estimation model. 
Despite 3-bit quantisation and 85\% unstructured pruning, the proposed compressed estimation model preserves accuracy close to the original BLS model~\cite{li2023broad}. 

\begin{table}[t]
    \centering
    \caption{Mean and maximum relative error of different harmonic estimation methods for each harmonic order}
    \label{tab:rel_error_methods}
    \begin{tabular}{lcccc|cccc}
        \hline
        \multirow{2}{*}{Method} & \multicolumn{4}{c|}{Mean relative error} & \multicolumn{4}{c}{Max relative error} \\
        \cline{2-9}
        & 1st & 3rd & 5th & 7th & 1st & 3rd & 5th & 7th \\
        \hline
        FFT   & 0.67 & 2.52 & 3.55 & 5.37 & 2.62 & 14.22 & 19.66 & 29.72 \\
        DWPT   & 0.67 & 2.57 & 3.84 & 4.62 & 2.62 & 14.51 & 20.41 & 33.36 \\
        MLP   & 0.83 & 0.80 & 0.82 & 0.79 & 2.00 & 2.00 & 2.02 & 2.00 \\
        RBF   & 0.53 & 0.50 & 0.56 & 0.48 & 1.24 & 1.09 & 1.29 & 1.10 \\
        AWN   & 0.50 & 0.50 & 0.53 & 0.48 & 1.02 & 1.03 & 1.09 & 1.01 \\
        BLS   & \textbf{0.49} & 0.50 & 0.52 & \textbf{0.48} & \textbf{1.00} & 1.01 & 1.01 & 1.00 \\
        \textbf{Prop.} & 0.51 & \textbf{0.48} & \textbf{0.51} & 0.50 & 1.09 & \textbf{0.97} & \textbf{1.00} & \textbf{0.99} \\
        \hline
    \end{tabular}
    \vspace{-1mm}
\end{table}

Table~\ref{tab:latency_methods_devices} compares the reported computational latency of different harmonic estimation methods and deployment platforms.
The latency reported for the proposed estimation models refers to the FPGA inference datapath after the input window is available.
As shown in the table, sparse optimisation reduces the inference latency of the proposed estimation model from 82\,ns for \textbf{Prop. (d)} to 70\,ns for \textbf{Prop. (s)}.
This low inference overhead supports real-time harmonic estimation updates in the control loop once each observation window has been acquired, making the proposed estimation model suitable for real-time harmonic reference generation.


\begin{table}[t!]
    \centering
    \caption{Latency comparison of different methods and platforms}
    \label{tab:latency_methods_devices} \scalebox{0.9}{
    \setlength{\tabcolsep}{5.5pt}
    \begin{tabular}{llr|llr}
        \hline
        Method & Platform & Latency & Method & Platform & Latency \\
        \hline
        GNO \cite{genccol2023efficient}  & F28027    & 0.31 ms          & FFT \cite{li2023broad}  & i5-9400F & 2.6 $\mu$s \\
        FFT \cite{li2024fast}            & ZYNQ7000 & 0.33 ms          & DWPT \cite{li2023broad} & i5-9400F & 10.4 ms \\
        DWPT \cite{tiwari2016hardware}            & AC-701     & 8.8 -- 13 $\mu$s          & MLP \cite{li2023broad}  & i5-9400F & 7 $\mu$s \\
        DWPT \cite{baraskar2022digital}  & AC-701    & 1.22 $\mu$s      & RBF \cite{li2023broad}  & i5-9400F & 22 $\mu$s \\
        \textbf{Prop. (d)}               & ZU7EV    & 82 ns            & AWN \cite{li2023broad}  & i5-9400F & 26.3 $\mu$s \\
        \textbf{Prop. (s)}               & ZU7EV    & \textbf{70 ns}   & BLS \cite{li2023broad}  & i5-9400F & 80.3 $\mu$s \\
        \hline
    \end{tabular}}
\end{table}

\begin{figure}[t]
    \centering
    \includegraphics[width=\columnwidth]{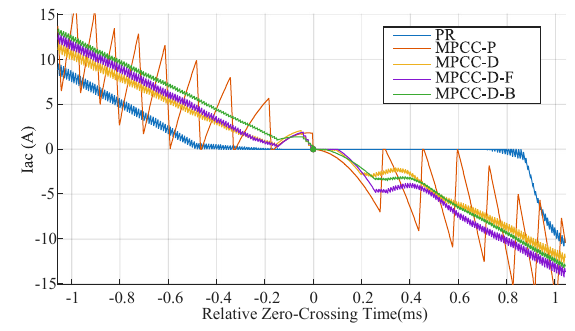}
    \caption{Zoomed view of the input current around the zero-crossing region under different control strategies.}S
    \label{fig:zero_crossing}
\end{figure}
\subsection{MPCC with Harmonic Estimator Online Evaluation}
In this section, we integrate our estimation model with the proposed duty-cycle predictive MPCC controller and evaluate them with a 7-kW EV OBC.
The current waveform around the grid-voltage zero-crossing region provides an intuitive indication of the control granularity, transient response, and dead-time-induced distortion under different control strategies. Therefore, it can be used to visually compare the current tracking quality and zero-crossing distortion suppression capability of different controllers.
In Fig.~\ref{fig:zero_crossing}, we compared the current with the control strategy PR, representing the Proportional Resonant control in~\cite{giroux2026evobc}, MPCC-P, MPCC with gate action prediction~\cite{ko2023model} and our duty cycle predictive MPCC with no harmonic reference as MPCC-D, FFT-based reference as MPCC-D-F and BLS-based prediction as MPCC-D-B.
The PR controller shows noticeable distortion in this region, since its current regulation is mainly determined by the resonant tracking loop and does not explicitly account for the converter switching model.
MPCC-P improves the current response by predicting the inductor current under candidate switch positions.
However, its binary control action limits the available current slopes, which can lead to local waveform irregularities when the required control effort lies between two switching states.
MPCC-D alleviates this problem by replacing the direct switching decision with a continuous duty-cycle command applied through fixed-frequency PWM.
When harmonic reference correction is further introduced, the controller compensates the estimated low-order harmonic components before the reference is tracked.

\begin{figure}[t]
    \centering
    \includegraphics[width=\columnwidth]{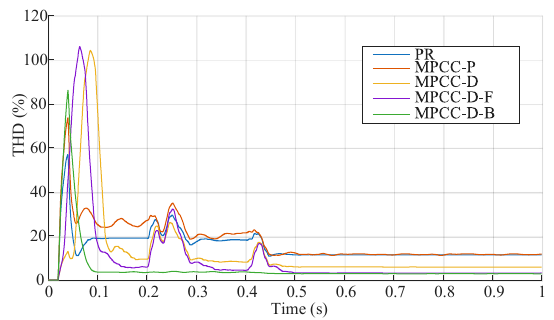}
    \caption{Dynamic evolution of input current $THD_i$ during EV charger start-up under different control methods.}
    \label{fig:thd_dynamic}
    \vspace{-3mm}
\end{figure}

We captured the system status within 1 second of OBC's operation, which covers both the OBC startup process and its stable operation.
Fig.~\ref{fig:thd_dynamic} shows the evolution of the $THD_i$ during the start-up process.
During OBC start-up, the DC-link state has not yet reached steady state. Since MPCC rapidly tracks the reference according to the prediction model and tends to adopt a relatively more aggressive control action, it may exhibit a higher $THD_i$ level than the conventional PR method during the transient start-up stage. After the system reaches steady state, the advantages of MPCC become more evident, as reflected by reduced oscillations and a significantly lower final steady-state $THD_i$ compared with the PR method.

\begin{table}[htbp]
\vspace{-3mm}
\centering
\caption{Performance comparison of different control methods}
\label{tab:control_comparison}
\begin{tabular}{lccccccc}
\toprule
\multirow{2}{*}{Method} & \multicolumn{5}{c}{RMS Amplitude} & $THD_i$ & Ripple \\
\cmidrule(lr){2-6}
 & 1 & 3 & 5 & 7 & 9 & (\%) & (\%) \\
\midrule
PR~\cite{giroux2026evobc}       & 30.53 & 2.67 & 1.43 & 1.07 & 0.81 & 11.77 & 5.95 \\
MPCC-P~\cite{ko2023model}   & 30.52 & 1.53 & 0.20 & 0.28 & 0.13 & 11.47 & 5.72 \\
MPCC-D   & 30.53 & 1.57 & 0.48 & 0.29 & 0.20 & 6.10  & 5.75 \\
MPCC-D-F & 30.51 & 0.75 & 0.26 & 0.18 & 0.21 & 3.37  & 5.60 \\
MPCC-D-B & 30.53 & 0.54 & 0.08 & 0.57 & 0.18 & 2.85  & 5.61 \\
\bottomrule
\end{tabular}
\end{table}

Table~\ref{tab:control_comparison} summarises the steady-state harmonic amplitudes, overall $THD_i$, and current ripple achieved by the different control methods. MPCC-D-F and MPCC-D-B achieve lower $THD_i$ and ripple levels because the estimated 3rd-, 5th-, and 7th-order harmonic components are incorporated into the current reference reconstruction process. Compared with FFT-based estimation, the BLS-based estimator provides more accurate harmonic references, resulting in more effective harmonic compensation. Although a given controller may not minimise every individual harmonic component simultaneously, the proposed harmonic-assisted duty-cycle MPCC achieves the lowest overall $THD_i$. The reference correction improves the overall harmonic profile of the input current rather than targeting a specific frequency component alone.
The experimental results of the two case studies demonstrate that the proposed duty-cycle predictive MPCC can selectively reduce harmonic contents and improve power quality when guided by the reference information provided by a more effective estimation model.
\section{Conclusion}\label{conclusion}
This paper presented a harmonic estimation-assisted duty cycle MPCC framework for the PFC stage of a single-phase AC/DC EV charger.
The proposed method integrates online harmonic estimation into the current reference generation path, enabling selective compensation of the dominant low-order harmonics.
By using duty cycle prediction with fixed frequency PWM, the controller provides smoother current regulation than finite control set MPCC, while the harmonic reference correction further improves the steady state current quality.
The proposed estimator achieves accurate and low-latency harmonic estimation on the edge side.
Closed-loop 7-kW OBC results show that the harmonic-assisted duty cycle predictive MPCC reduces the steady state current THD from \(6.10\%\) under duty cycle MPCC to \(2.85\%\), with no evident increase in current ripple.
Future work will focus on validating the proposed method on the real main circuit and exploring edge reinforcement learning for further harmonic optimisation.

\bibliographystyle{IEEEtran}
\bibliography{references}

\end{document}